\documentclass[fleqn,usenatbib,useAMS]{mnras}

\usepackage{graphicx}	
\usepackage{amsmath}	
\usepackage{amssymb}	
\usepackage{multicol}        
\usepackage{bm}		
\usepackage{pdflscape}	
\usepackage{comment}

\defcitealias{2011ApJ...741...99C}{Cen+11}

\usepackage[T1]{fontenc}
\usepackage{ae,aecompl}

\usepackage{newtxtext,newtxmath}

\title[MNRAS]{The Fates of Merging Supermassive Black Holes and a Proposal for a New Class of X-Ray Sources}

\author[]{
Charles Zivancev$^{1}$\thanks{Contact e-mail: \href{mailto:csz2104@columbia.edu}{csz2104@columbia.edu}},
Jeremiah Ostriker$^{1}$,
Andreas H.W. K\"upper$^{2,3}$
\\
$^{1}$Department of Astronomy, Columbia University, 550 West 120th Street, New York, NY 10027, USA\\
$^{2}$QuantCo, Inc., 955 Massachusetts Avenue, Cambridge, MA 02139, USA\\
$^{3}$Hubble Fellow\\
}
\date{Last updated \today; in original form \today}

\pubyear{2019}

\begin{document}
\label{firstpage}
\pagerange{\pageref{firstpage}--\pageref{lastpage}}
\maketitle

\begin{abstract}
We perform N-body simulations on some of the most massive galaxies extracted from a cosmological simulation of hierarchical structure formation with total masses in the range $10^{12} M_{\sun} < M_{tot} < 3\times 10^{13} M_{\sun}$ from $4\geq z \geq 0$.  After galactic mergers, we track the dynamical evolution of the infalling black holes (BHs) around their host's central BHs.  From 11 different simulations, we find that, of the 86 infalling BHs with masses > $10^4 M_{\sun}$, 36 merge with their host's central BH, 13 are ejected from their host galaxy, and 37 are still orbiting at $z=0$.  Across all galaxies, 33 BHs are kicked to a higher orbit after close interactions with the central BH binary or multiple, after which only one of them merged with their hosts.  These orbiting BHs should be detectable by their anomalous (not Low Mass X-ray Binary) spectra.  The X-ray luminosities of the orbiting massive BHs at $z=0$ are in the range $10^{28}-10^{43}$ $\mathrm{erg}~\mathrm{s}^{-1}$, with a currently undetectable median value of $10^{33}$ $\mathrm{erg}~\mathrm{s}^{-1}$.  However, the most luminous $\sim$5\% should be detectable by existing X-ray facilities.
\end{abstract}


\section{Introduction}\label{sec:introduction}
Simulations of structure formation in a $\Lambda$CDM universe show us how the massive galaxies we see today were formed through a myriad of minor and major mergers between (proto) galaxies. As most galaxies are believed to contain central massive black holes even at high redshifts (\citet{1998AJ....115.2285M}, \citet{2003ApJ...589L..21M}, \citet{2004ApJ...604L..89H}, \citet{2013ARA&A..51..511K}), each of these mergers set the stage for a potential encounter between massive black holes. Here we investigate what happened to those black holes over a timeframe of billions of years.

What are possible outcomes for massive black holes during galaxy mergers?  There are three possibilities: they merge or form a binary with the central massive black hole of their new host galaxy, they get ejected through three-body interactions with an existing central black-hole binary, or they remain in orbit in the merged galaxy.  Combinations of these options are also possible with, e.g., the (spin-induced) kick at merger leading to an ejection or extended orbit of the merged black hole. We will argue in this paper that a very common outcome -- especially for lower-mass black holes -- is for the injected BHs to remain as orbiting X-ray sources in their host galaxies, waiting to be identified by current observational techniques.

Orbiting black holes have been studied in the past using both semi-analytical methods and dynamical simulations.  \citet{2003ApJ...582..559V} followed the merger history of dark matter haloes and associated BH mergers using Monte-Carlo realizations, and found a population of BHs orbiting in galaxy haloes and the intergalactic medium at the present epoch.  In a follow-up study, \citet{2005MNRAS.358..913V} found that the orbiting black holes should have luminosities in X-ray $\lessapprox10^{39} M_{\sun}$.  \citet{2008MNRAS.390.1311B} looked at the effects of gravitational recoil due to BH mergers on the growth of SMBHs, and found that the recoiled BHs can orbit their host galaxy for $10^6-10^9$ years before returning to the center, while accreting gas equivalent to approximate $10{\%}$ of their initial mass.  \citet{2017MNRAS.472.1526C} simulated recoiling black hole trajectories, altering various parameters in their simulations and observing the effects on escape velocities.  They found that the overall mass of the host halo and the accretion rate onto the halo are the largest factors affecting the escape velocity of a recoiled BH.  In a large-scale cosomological simulation (ROMULUS25) of Milky Way-mass galaxies, \citet{2018ApJ...857L..22T} found that such haloes would have approximately 5 SMBHs within $10kpc$ from the galactic center, and 12 SMBHs within they virial radius.  \citet{2020MNRAS.495.4681I} used a semi-analytical model to study the mass assembly and spin evolution of SMBHs, as well as the effects of gravitational recoil on central and orbiting black holes.  One of their findings is that the number and masses of orbiting black holes corresponds with the halo and stellar mass of the host galaxy, with their location within the galaxy being approximately $0.5R_{200}$.

But how common are encounters of massive black holes?  For massive galaxies, we know that minor mergers are frequent from the observed evolution of the size and mass of these systems \citep{2010ApJ...718L..73V, 2008ApJ...677L...5V, 2019MNRAS.484..595M}. However, if we assume that all galaxy mergers lead to SMBH mergers we end up overpredicting the  gravitational wave background, as constrained by observations from pulsar timing arrays (PTAs, see \citealt{2008MNRAS.390..192S, 2009MNRAS.394.2255S, 2013MNRAS.433L...1S, 2014ApJ...789..156M, 2015ApJ...799..178K, 2016APS..APRR18003T, 2018ApJ...856...42S, 2018MNRAS.479.4017I, 2018ApJ...863L..36I, 2018NatCo...9..573M}). In fact, dynamical studies of black hole encounters have shown that most encounters between two supermassive black holes will result in a stalling binary, unable to get close enough for gravitational wave emission to drive the merger \citep{1980Natur.287..307B, 2003AIPC..686..201M}.

However, we do not often see SMBH binaries at the centers of massive ellipticals (but see \citealt{2016MNRAS.463.2145C}).  So, what does happen?  \cite{2018MNRAS.473.3410R} indicate that dynamical interactions among multiple orbiting black holes, which will eject a non-negligible fraction of the mass, may solve this problem.  The present paper also addresses this purported dynamical solution, focusing attention on the large fraction of lower-mass black holes that remain to be detected as they orbit in massive galaxies.  We argue that, as a result of few-body interactions among the resident and infalling black holes, some are ejected, some remain in extended orbits, and some (the few most massive ones) do in fact merge. The merger rates we find do not exceed the PTA limits, and some of the orbiting BHs that our simulations predict to exist may actually be detectable via their ability to accrete gas and emit radiation.

For this study, we look at the merger history of 11 exemplary galaxies across the galaxy mass spectrum extracted from a cosmological simulation of hierarchical structure formation. We investigate how, after merging with incoming galaxies, SMBHs sink into the cores of the hosts and interact with the resident black hole. We show that gravitational interactions of multiple SMBHs are most probable in high-mass galaxies with total mass $10^{12} M_{\sun} < M < 3\times10^{13} M_{\sun}$. Galaxies with lower masses have too few mergers with SMBH hosting galaxies. Galaxies with higher masses are more extended, making dynamical friction processes less efficient and hence failing to drive SMBHs into the host galaxy core.  \cite{2012MNRAS.422.1306K} and \cite{2018MNRAS.473.3410R} performed similar N-body simulations of merging galaxies and black holes, although their haloes were from dark matter-only merger trees from the Millennium Simulation \citep{2005Natur.435..629S}, which have very different merger histories compared to the hydrodynamical simulations in this study.  Additionally, most of the halo masses in the \cite{2012MNRAS.422.1306K} simulations were of mass $M_{halo} > 10^{14} M_{\odot}$.

This paper is organized as follows: in Section\,\ref{sec:methods} we describe the cosmological simulations from which we use the merger history to set up our idealized numerical simulations. We present the few-body integration code, \textsc{AR-Chain} that we used for our simulations of SMBH dynamics, and the modifications we made to this code in order to deal with a host galaxy's gravitational potential. In Section\,\ref{sec:results}, we show the results of our 11 exemplary simulations of galaxies growing with time and acquiring new SMBHs. We analyze how the SMBHs spiral into the core of their new host galaxies due to dynamical friction, and how interactions with the host black hole and other orbiting BHs leads to near-ejections or mergers. We then estimate the X-ray luminosities expected of the orbiting black holes.  The final Section\,\ref{sec:conclusions} contains a discussion of the results and our conclusions.

\section{Methods}\label{sec:methods}

\subsection{Overview of simulation}
The computational challenge of this investigation is a wide range of time-scales. Galactic evolution and galactic orbits take hundreds of millions to billions of years, while orbital times of BH binaries may require time steps of years for accurate integration. We overcome this challenge by simplifying the model where possible, and by focussing on the main drivers of BH dynamics: the galactic background potential, dynamical friction, and strong BH-BH encounters.

Our simulations focus on elliptical galaxies with central SMBHs. The galaxies were given a dark matter background potential using the Stone-Ostriker profile \citep{2015ApJ...806L..28S}, which is a three-parameter potential-density pair, whose quantities such as density, potential, and binding energy can be written in closed form, having a profile similar to that of a truncated isothermal sphere.  The galaxies were evolved from $4\geq z \geq 0$.  Orbiting black holes were periodically introduced into the "host" galaxy following mergers and their dynamical interaction with the background potential and central SMBH were followed, as described further below.

For the numerical simulations presented here, we used a modified version of the algorithmic chain integrator \textsc{AR-Chain} developed by \citet{2006MNRAS.372..219M}. It uses algorithmic chain regularization for high-precision integration of few-body dynamics, and is capable of handling velocity-dependent forces efficiently. It includes relativistic post-Newtonian terms up to order 2.5 \citep{2008AJ....135.2398M}.

\subsection{Merger tree Data}
The merger tree data we used as the input to our simulations is the result of previous work by \citet{2011ApJ...741...99C, 2011ApJ...742L..33C, 2012ApJ...753...17C, 2012ApJ...748..121C, 2013ApJ...770..139C}  (known as the LAOZI simulations \citealt{2014ApJ...781...38C}), as well as  \citet{2012MNRAS.425..641L}, and \citet{2015ApJ...799..178K}.  A brief synopsis of the work that led to our input data is described below.

\citet{2011ApJ...741...99C} studied the "cosmic downsizing" effect (e.g., \citealt{1996AJ....112..839C}) using high-resolution large-scale hydrodynamic galaxy formation simulations.  His work produced physical parameters for galaxies such as position, velocity, total mass, stellar mass, gas mass, mean formation time, mean stellar metallicity, mean gas metallicity, star formation rate, luminosities, etc.

\citet{2012MNRAS.425..641L}, using \citet{2011ApJ...741...99C}'s work as a basis, created galactic merger trees that compared the properties of in-situ and accreted stellar mass of galaxies at redshift snapshots between $4\geq z \geq0$.

\citet{2015ApJ...799..178K} furthered the merger tree data produced by \citet{2012MNRAS.425..641L} by estimating the evolution of the SMBH population.  In their simulations, they seed a black hole in any galaxy that reaches their prescribed lower mass limit ($M_* \simeq 10^8 M_{\sun}$).  The initial mass of the seeded black hole is then taken to be $nM_{*, bulge}$, where $n$ is drawn from a log-normal distribution with median $\simeq10^{-3}$ and intrinsic scatter of 0.35.  The black holes subsequently grow through BH-BH mergers and gas accretion.  The accretion component is proportional to the stellar mass formed in the host galaxy in a given redshift slice times the bulge fraction of the host galaxy.  They assume growth of SMBHs to be proportional to that of the bulge stellar mass of the galaxies.  They further assume a constant relation between bulge and total stellar mass at all redshifts, although some observations suggest that the bulge mass fraction for a fixed stellar mass should increase with decreasing redshift (\citealt{2012MNRAS.423.1992S, 2013MNRAS.428.1351G, 2014MNRAS.441..417A}), with the result that their SMBHs grow faster at high redshift than would be expected in reality.

\citet{2015ApJ...799..178K} also had a prescription for re-seeding a galaxy with a central black hole in the event the previous one was ejected due to gravitational recoils. If the galaxy subsequently produced an additional $10^9M_{\sun}$ in unscaled galactic stellar mass, they re-seed the galaxy using the same method described above.  In our simulations, the central black holes were never ejected, and we did not need to use the above-described re-seeding.

The primary data we extracted from the above work was organized as follows:
\begin{itemize}
    \item Galactic properties - Each galaxy was distinguished by an id number, with properties such as stellar mass, dark matter (DM) mass, central black hole id number, and orbiting black hole id numbers, recorded at 37 redshift slices between $4\geq z \geq0$. The original \citet{2015ApJ...799..178K} simulations had 38 redshift slices.  However, one was dropped due to an error in the dataset.
    \item Black Hole properties - Each black hole was also distinguished by an id number, with properties such as seed mass, accreted mass, its host galaxy id number, and time after $z=4$ at which the black hole entered its host galaxy, if it was not the central black hole.
\end{itemize}

The merger tree data from \citet{2012MNRAS.425..641L} and \citet{2015ApJ...799..178K} provided 1,830 galaxies in total.  However, not all the galaxies were suitable for our simulations.  We placed further requirements as follows:
\begin{itemize}
\item The galaxies had to exist through the entire simulation ($4\geq\;z\;\geq 0$).  If they merged with other galaxies, they had to have been the "surviving" galaxy at each merger.
\item They had to have accumulated orbiting black holes by $z = 0$.
\end{itemize}

The vast majority of the full set of galaxies (1,537) were eliminated because they did not come into existence until after $z=4$.  Of the remaining 293 galaxies, 67 of them existed at $z=0$, and 51 of them also had black holes orbiting at $z=0$.  In considering time and resources, we chose the 11 galaxies that had the largest number of black holes to run the simulations.  The black holes in these galaxies accounted for nearly $84{\%}$ of all orbiting black holes across all the galaxies, so it represented a reasonable balance between completeness and efficiency.

The following cosmological parameters were used in our simulations, consistent with both \citet{2012MNRAS.425..641L} and \citet{2015ApJ...799..178K}:   $\Omega_M = 0.28$, $\Omega_b = 0.046$, $\Omega_\Lambda = 0.72$, $\sigma_8 = 0.82$, $H_0 = 100h^{-1}\mathrm{Mpc}^{-1} = 70 \mathrm{km}\;\mathrm{s}^{-1} \mathrm{Mpc}^{-1}$, and $n = 0.96$.

\subsubsection{Stellar Mass Adjustment}
As we will explain further in Section \ref{Galaxy background potential}, in our simulations orbiting black holes are inserted into the host galaxy at the effective radius, $R_e$, which is partially dependent on stellar mass, $M_*$ (see Equation \ref{re}).  For every doubling of the stellar mass, $R_e$ increases by a factor of 1.66.  Additionally, the velocity dispersion at the effective radius, $\sigma(R_e)$, which is used to set the parameter $r_h$ (see Equation \ref{jerry}) of our dark matter profile, is also dependent on $M_*$.  In this case, a doubling of $M_*$ would increase $\sigma(R_e)$ by a factor of 1.15.  These variations have direct effects on the ability of the orbiting black holes to merge with the central black hole, and thus can affect our merger statistics.  Therefore, it is important to estimate our stellar masses with reasonable accuracy.

In cosmological simulations, it is common to overproduce stellar mass, with a general over-efficiency in the range of 2-4 times those measured in observations \citep{1996ApJS..105...19K, 2010MNRAS.404.1111G, 2010ApJ...725.2312O}.  One cited reason for this overproduction is a lack or underestimation of supernova feedback effects in the form of thermal and/or kinetic energy.  However, even when thermal energy from supernova feedback is considered, the effects may be reduced due to the energy being quickly radiated away by rapidly cooling surrounding gas \citep{1996ApJS..105...19K}.  A second common reason for overproduction of stars in cosmological simulations is a lack of modeling for AGN feedback, which may be responsible for solving the phenomenon called the "cooling flow paradox" \citep{2001MNRAS.321L..20F} that suppresses star formation. 

\citet{2012MNRAS.425..641L}, which is the source of our stellar masses, noted that the efficiency of star formation in their simulations, defined as $f_\mathrm{bary}=M_*/M_{\mathrm{DM}}(\Omega_\mathrm{DM}/\Omega_\mathrm{b})$ (where $M_\mathrm{DM}$ is the mass of the dark matter halo), was approximately 0.6.  Compared with the expected range of $0.10 \loa f_\mathrm{bary} \loa 0.15$ that they referenced from \citet{2012ApJ...746...95L}, their stellar masses were a factor of roughly 4 times greater.

Given the overestimate of stellar mass in the merger tree data, we rescaled our stellar masses in order to match the observational and abundance matching data from \citet{2018AstL...44....8K}. In their study, they estimated star formation efficiency and the stellar mass-halo mass relation in massive haloes.  They obtained their data from two sources: a study of nine nearby clusters ($z<0.1$) using Chandra X-ray observations \citep{2009ApJ...692.1033V}, and one of 12 cluters from \citet{2013ApJ...778...14G}.

\begin{figure}
\includegraphics[width=1.0\columnwidth]{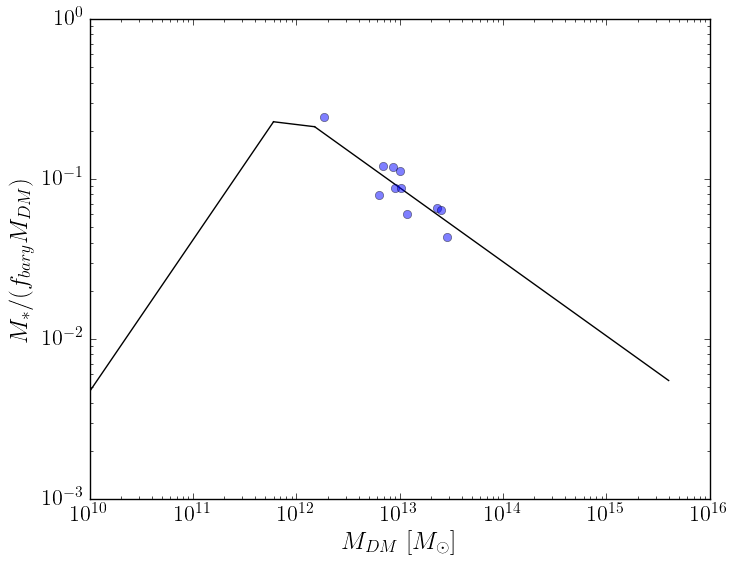}
\caption{Approximation of Figure 11 from \citet{2018AstL...44....8K}.}  $f_\mathrm{bary}$ is the universal baryon fraction as a function of total halo mass.  Purple dots shown are examples of our rescaled stellar masses.  We rescaled our stellar masses at $z=0$ by using a lognormal distribution.  For a given galaxy's dark matter mass, the mean of the distribution was taken to be the point on the abundance matching line at that dark matter mass.  The standard deviation of that mean was taken arbitrarily to be 0.3.
\label{fig:stellar1}
\end{figure}

Figure \ref{fig:stellar1} is a partial, approximate, replication of Figure 11 from \citet{2018AstL...44....8K} (abundance matching line).  We rescaled our stellar masses at $z=0$ by using a lognormal distribution.  For a given galaxy's dark matter mass, the mean of the distribution, $\mu$, was taken to be the point on the abundance matching line at that dark matter mass.  The standard deviation was assumed arbitrarily to be $0.3\mu$.  After rescaling the stellar masses at $z=0$, we used that proportionality to rescale the masses through the entire simulation from $4\geq z \geq0$.  Table \ref{table:gal_char} shows the original and rescaled stellar masses.

\begin{table}
\begin{center}
\caption{Galaxy characteristics at $z=0$. $M_\mathrm{DM}$ is dark matter mass.  $M_\mathrm{CBH}$ is central black hole mass.  Units are $M_{\odot}$.}
\begin{tabular} {|c | c | c| c|}
\hline
$M_\mathrm{DM}$ & $M_\mathrm{CBH}$ & $M_{*} \mathrm{(orig.)}$ & $M_{*} \mathrm{(rescaled)}$ \\
\hline
 2.85 $\times 10^{13}$ 	&	 1.46 $\times 10^9$	&	5.58 $\times 10^{12}$ 	&	3.30 $\times 10^{11}$  \\
 2.48 $\times 10^{13}$ 	&	 2.40 $\times 10^9$	&	4.80 $\times 10^{12}$ 	&	2.25 $\times 10^{11}$  \\
 2.29 $\times 10^{13}$ 	&	 1.22 $\times 10^9$	&	4.46 $\times 10^{12}$ 	&	2.68 $\times 10^{11}$  \\
 1.18 $\times 10^{13}$ 	&	 3.62 $\times 10^9$	&	2.32 $\times 10^{12}$ 	&	2.83 $\times 10^{11}$  \\
 1.03 $\times 10^{13}$ 	&	 9.85 $\times 10^9$	&	1.96 $\times 10^{12}$ 	&	1.62 $\times 10^{11}$  \\
 1.00 $\times 10^{13}$ 	&	 9.32 $\times 10^9$	&	1.95 $\times 10^{12}$ 	&	2.49 $\times 10^{11}$  \\
 8.92 $\times 10^{12}$ 	&	 5.60 $\times 10^9$	&	1.74 $\times 10^{12}$ 	&	3.20 $\times 10^{11}$  \\
 8.56 $\times 10^{12}$ 	&	 6.86 $\times 10^9$	&	1.62 $\times 10^{12}$ 	&	1.83 $\times 10^{11}$  \\
 6.94 $\times 10^{12}$ 	&	 2.02 $\times 10^9$	&	1.72 $\times 10^{12}$ 	&	1.51 $\times 10^{11}$  \\
 6.29 $\times 10^{12}$ 	&	 1.22 $\times 10^9$	&	1.23 $\times 10^{12}$ 	&	1.87 $\times 10^{11}$  \\
 1.84 $\times 10^{12}$ 	&	 2.21 $\times 10^9$	&	5.52 $\times 10^{11}$ 	&	6.42 $\times 10^{10}$  \\
\hline
\end{tabular}
\end{center}
\label{table:gal_char}
\end{table}

\subsection{Simulation Physics}
In this section, we describe the underlying physics of our simulations, including choice of galaxy background potential, BH diffusion through the galaxy, and gravitational wave kicks as a result of black hole mergers.

\subsubsection{Galaxy background potential} \label{Galaxy background potential}
To model the dark matter halo distribution in our simulations, we used the Stone-Ostriker profile \citep{2015ApJ...806L..28S}, which is a three-parameter potential-density pair.  Quantities such as density, potential, and binding energy can be written in closed form, which speeds up computations.  It is essentially an analytic form of a finite, cored isothermal mass distribution:
\begin{equation} \label{jerry}
\rho(r) = \frac{\rho_c}{(1+r^2/r_{c}^2)(1+r^2/r_{h}^2)}.
\end{equation}
Here $\rho_c$ is the central density, $r_c$ is the core radius, and $r_h$ is the outer halo radius.

The central density, $\rho_c$, can be found from inverting Equation 5 in  \citet{2015ApJ...806L..28S} for the galaxy's total mass:

\begin{equation} \label{rhoc}
M_{DM} = \frac{2\pi^2r_{c}^2r_{h}^2\rho_c}{r_h+r_c}.
\end{equation}

In order to parameterize $r_c$ and $r_h$, we can utilize the relationship between the distribution of light and stellar velocity dispersion at the effective radius, and the shape of the dark matter profile.  The effective radius $R_{e}$ and velocity dispersion at the effective radius $\sigma(R_e)$, respectively, can be obtained from (\citealt{2009ApJ...706L..86N, 2010ApJ...725.2312O, 2014ApJ...789..156M}):

\begin{equation} \label{re}
R_{e} = 2.5 \mathrm{kpc}\left(\frac{M_*}{10^{11}M_{\odot}}\right)^{0.73}(1+z)^{-0.98},
\end{equation}
\begin{equation} \label{sig}
\sigma(R_{e}) = 190\mathrm{km}\;\mathrm{s}^{-1}\left(\frac{M_{*}}{10^{11}M_{\odot}}\right)^{0.2}(1+z)^{0.47}.
\end{equation}

We also give $r_c$ an initial value of 100 pc, which is reasonable for a cored massive system.  We can then equate the value for $\sigma({R_e})$ obtained from Equation \ref{sig} to the analytic expressions for $\sigma$ in the Stone-Ostriker profile (cf. Equations 9 and A1-A4 in \citealt{2015ApJ...806L..28S}).  The only unknown is $r_h$, which we can solve for using a simple recursive Newton method.  Whether $\sigma_{near}$ or $\sigma_{far}$ is used from \cite{2015ApJ...806L..28S} is determined by whether $R_e$ is less than or greater than $\sqrt{r_c r_h}$.

At each timestep in the code, $r_h$ is updated according to the stellar mass and Equations \ref{re} and \ref{sig}.  The core radius, $r_c$, is recalculated only if an orbiting black hole enters within $r_c$.  The work done by diffusion, as described in Section \ref{psd} is calculated, the total potential energy and dynamical friction are updated, and $r_c$ is solved for from Equation 8 in \citet{2015ApJ...806L..28S} based on conservation of energy.

\subsubsection{Phase-space diffusion} \label{psd}
Much of this subsection was borrowed from \citet{2017MNRAS.467.4180S}, Section E1, which made use of the same modified version of \textsc{AR-Chain}.

Weak encounters with background stars and dark-matter particles will let the SMBHs diffuse through phase space while they are orbiting within the gravitational potential of the galaxy. The diffusion can be expressed as change in velocity of an SMBH by $\Delta \vec{v}$ per unit time. We can split this change into a component along the direction of motion of the SMBH, and one perpendicular to that. Following \citet{2008gady.book.....B}, the diffusion coefficients can be expressed as 
\begin{eqnarray}\label{eq:df}
D[\Delta v_\parallel] & = & -\frac{4\pi G^2\rho(r)M_{BH}\ln\Lambda}{\sigma^2}f(\chi),\\
D[(\Delta v_\parallel)^2] & = & \frac{4\sqrt{2}\pi G^2\rho(r)M_{BH}\ln\Lambda}{\sigma}\frac{f(\chi)}{\chi},\\
D[(\Delta \vec{v}_\bot)^2] & = & \frac{4\sqrt{2}\pi G^2\rho(r)M_{BH}\ln\Lambda}{\sigma}\left[\frac{\mbox{erf}(\chi)-f(\chi)}{\chi}\right],
\end{eqnarray} 
where $\Delta v_\parallel \equiv \Delta \vec{v}\cdot\vec{v}/v$ is the velocity change in direction of motion, and $\Delta \vec{v}_\bot \equiv \Delta \vec{v} - \Delta v_\parallel \cdot\vec{v}/v$ is the velocity change perpendicular to the direction of motion. Here, $\sigma$ is the local velocity dispersion of the host galaxy, $M_{BH}$ is the mass of the black hole, and $\chi = \frac{v}{\sqrt{2}\sigma(r)}$. The function $f(\chi)$ is given by 
\begin{equation}
f(\chi) \equiv \frac{1}{2\chi^2}\left(\mbox{erf}(\chi)-\frac{2\chi}{\sqrt{\pi}}\exp\left(-\chi^2\right)\right).
\end{equation}
We approximate the factor $\Lambda$ in the Coulomb logarithm as
\begin{equation}
\Lambda \equiv \left(\frac{M_{M_{DM}}}{M_{BH}}\right)\left(\frac{r}{r_h}\right).
\end{equation}
We can identify Equation \ref{eq:df} as the dynamical friction term. The second term introduces a variance of the friction term, and even allows the BHs to be accelerated when the velocity of a BH is sufficiently small. The third term introduces a change in velocity perpendicular to the direction of motion of the BH. It is a randomly oriented vector, and hence causes the BHs to depart from smooth orbits due to the small random velocity kicks. The last two terms will establish that the BHs are ultimately in energy equipartition with the background stars.
The velocity changes $\Delta v_\parallel$ and $\Delta\vec{v}_\bot$ per unit time $\Delta t$ can be computed with the above equations. Both changes are normally distributed, where the mean, $\mu$, and the variance, $\Sigma$, of the distributions are given by
\begin{eqnarray}
\mu_\parallel &=& D[\Delta v_\parallel]\Delta t,\\
\Sigma_\parallel &=& D[(\Delta v_\parallel)^2]\Delta t,\\
\mu_\bot &=& 0,\\
\Sigma_\bot &=& D[(\Delta \vec{v}_\bot)^2]\Delta t.
\end{eqnarray}
We compute the diffusion coefficients for each black hole at each time step, and modify its velocity on a Monte Carlo basis. For each time step we draw a random orientation before adding the perpendicular velocity change to the respective BH. Hence, the BH's modified velocity, $v_f$, is computed using
\begin{eqnarray}
\vec{v}_f &=& \vec{v}_0 + \Delta v_\parallel \hat{v}_\parallel + \Delta v_\bot \hat{v}_\bot,\\
\Delta v_\parallel &=& \mathcal{N}(\mu_\parallel, \Sigma_\parallel),\\
\Delta v_\bot &=& \mathcal{N}(\mu_\bot, \Sigma_\bot).
\end{eqnarray}
The change of energy, $\mbox{d}E_{BH}$, of the orbiting black hole due to phase-space diffusion is given back to the galactic background potential, with $\mbox{d}E = -\mbox{d}E_{BH}$. As a consequence of this energy transfer, inspiralling black holes will cause an expansion of the galactic mass distribution. For this purpose we calculate the change in potential energy, $\mbox{d}W$, of the host galaxy using
\begin{eqnarray}
E &=& T + W = \frac{1}{2}W,\\
\mbox{d}W &=& -2\,\mbox{d}E_{BH},
\end{eqnarray}
where we made use of the virial theorem $2T+W =0$. With this change in potential energy we can calculate a new radius for the galactic background potential at each integration step. For simplicity, we assumed that the relatively low mass of the infalling black holes compared to the core mass of the host galaxies would only affect the size of the core radius, and hence ignored the feedback effect when the black holes were outside of the core radius.

\subsubsection{Gravitational wave recoils and escape}
The code \textsc{AR-Chain} (\citet{2006MNRAS.372..219M} and \citet{2008AJ....135.2398M}) includes post-Newtonian terms up to order 2.5. The SMBHs can therefore merge via gravitational wave emission. We include gravitational wave recoils (\citet{2007ApJ...659L...5C}, \citet{2011PhRvD..83b4003L}, \citet{2013PhRvD..87h4027L}, \citet{2012ApJ...744...45B}) following the prescription outlined in \citet{2015ApJ...799..178K}, which is based on the fitting formula by \citet{2012PhRvD..85h4015L}.  We assume that a merger will be inevitable when the separation between two SMBHs becomes smaller than four times the Schwarzschild radius of the more massive black hole.  At the moment of the merger, we also assume that the spin vectors of the two SMBHs are randomly aligned.  There are reasons to believe that the actual spins of the binary BHs may not be randomly aligned, as described further below.

A result of the anisotropic emission of gravitational waves is that the merged binary often experiences a linear kick in the opposite direction of the emission of the GW due to conservation of linear momentum.  This kick can result in velocities of the merged binary from approximately $200-5000\ km\ s^{-1}$ (see \citealt{2007PhRvL..98w1101G, 2007PhRvL..98i1101G, 2007PhRvL..98w1102C, 2011PhRvL.107w1102L}).  In our simulations, the binaries never stray more than a few tens of parsecs from the nucleus of the host galaxies.  As mentioned above, we calculate the spins and kicks (recoil velocities) of the black holes in the same way that \citet{2015ApJ...799..178K} implemented it (their Equations 11), which in turn followed the implementation from \citet{2012PhRvD..85h4015L}.  We also adopt a value of $\xi$ = 145$^{\circ}$ based on the numerical results for quasi-circular merger configurations from \citet{2008PhRvD..77d4028L}.

The kick distribution is approximated in the way described above, but the actual distribution of kicks may vary since, for example, the alignment of BH spins may not be fully random, as demonstrated in \citet{2012PhRvD..85h4015L} and \citet{2010MNRAS.402..682D}. Accretion disks surrounding the merging black holes may help align spins between the merging black holes on a timescale shorter than the time needed to merge.

Black holes can also eject each other via strong three-body interactions. We remove SMBHs from the simulations once they move beyond $r_h$, assuming that it will take them more than a Hubble time to find their way back into the center of the host galaxy, or that they have achieved escape velocity.

\subsection{Simulation setup}\label{subsec:simset}
For each galaxy in the simulations, we only used black holes that would theoretically have merged with the center of the host galaxy within a dynamical friction time, $t_{fric}$, of less than 100 times a Hubble time, with $t_{fric}$ defined by the Chandrasekhar formula from \citet{2008gady.book.....B} as:
\begin{equation}\label{tfric}
    t_{fric} = \frac{19}{6}\left(\frac{R_\mathrm{e}}{5\mathrm{kpc}}\right)^2\frac{\sigma(R_\mathrm{e})}{200\mathrm{km}\,\mathrm{s}^{-1}}\frac{10^8\,M_{\odot}}{M_\mathrm{BH}} \text{  Gyr},
\end{equation}
The formula above was not a factor during the simulations.  We used the full dynamical friction model (section \ref{psd}) during the simulations.  The orbiting black holes were injected into their host galaxy at redshift $z$, at a distance from the galactic center of $R_{e}$ (Equation \ref{re}) following galaxy mergers.  Their initial velocity was arbitrarily chosen to be the circular velocity at that radius, $v_c = \sqrt{GM(R_e)/R_e}$, with orientation, $v_x$, $v_y$, and $v_z$, randomly chosen.

\section{Simulation Results}\label{sec:results}\label{sec:results_overview}
Here we review the primary results of our simulations, focusing on the various interactions of the orbiting black holes with their host SMBHs. In the next section, we will investigate the luminosity characteristics of the orbiting black holes that did not merge at $z=0$.

In this section, "orbiting" black holes refer to all black holes that are still orbiting within their host galaxy at $z=0$.  "Kicked" black holes are a subset of orbiting black holes, in that they are still orbiting within their host galaxy, but at some point during the simulation were kicked to a higher orbit due to three-body scattering.  "Ejected" black holes are those that were given a large enough of a recoil velocity due to three-body scattering that they were ejected from their host galaxy.  To clarify, any discussion of kicked black holes below only refers to those that are still orbiting their host galaxy.  Similarly, discussions of ejected black holes only refer to those that were ejected.

\begin{figure}
\begin{center}
\includegraphics[width=1.0\columnwidth]{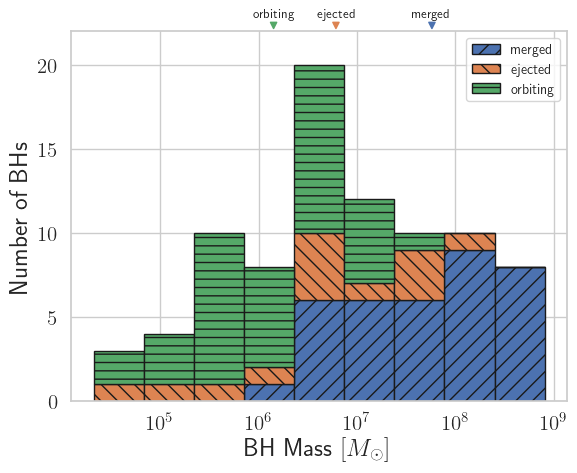}
\caption{Histogram of merged, ejected, and orbiting SMBHs at $z=0$ as a function of BH mass.  Ejected black holes span almost the entire mass range, while lower-mass black holes are orbiting and higher mass black holes tended to merge.  Colored arrows along upper x-axis represent median masses of respective sets.  Histogram bars are stacked.}
\label{fig:meosmbh}
\end{center}
\end{figure}

Figure \ref{fig:meosmbh} shows the distribution of merged, ejected and orbiting black holes (stacked) across all galaxy simulations as a function of the mass of the infalling black holes.

Of the 86 SMBHs used for the simulations, 36 black holes merged with their respective host black hole within a Hubble time, 37 remained orbiting at the end of the simulations, and 13 were ejected from their galaxies as a result of sufficiently large kicks due to three-body interactions with the host and other black holes.

Given the log distribution of masses, one can see the general trend of larger mass BHs ($\goa\, 10^7\, M_{\odot}$) tending to merge with their respective host BH, while lower mass BHs continue to orbit at $z=0$.  However, there is some overlap with some BHs greater than $10^7\, M_{\odot}$ still orbiting and those less than $10^7\, M_{\odot}$ having merged.

\subsection{Mergers}\label{sec:mergers}
\begin{figure}
\begin{center}
\includegraphics[width=1.0\columnwidth]{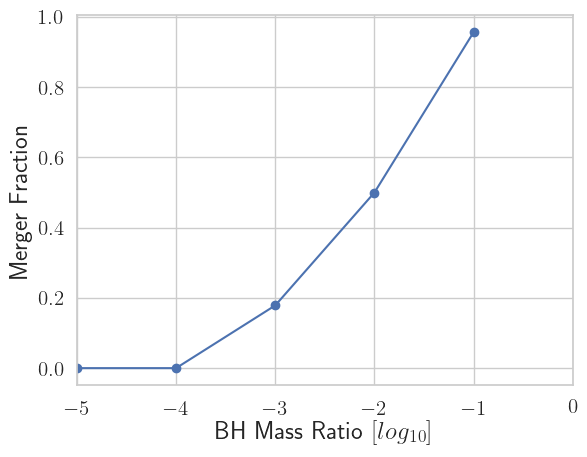}
\caption{Fraction of BHs that merge with the central BH as a function of the BH mass ratio.  Each point represents the fraction of galaxy mergers that led to BH mergers for each BH merger ratio bin.  Note that a "bin" at each point is comprised of the mass ratio at that point up to, but not including, the next point on the x-axis.}
\label{fig:q_merge_fraction}
\end{center}
\end{figure}

The tendency of larger mass black holes to merge with their host can perhaps be more clearly seen by looking at the merger mass ratio between the incoming black hole and the host.  Figure \ref{fig:q_merge_fraction} displays the fraction of galaxy mergers that result in black hole mergers versus the merger mass ratio, defined as $m_2/m_1$, where $m_2$ is the orbiting black hole and $m_1$ is the host black hole.  The plot monotonically increases with increasing mass ratio, reaching nearly unity for mass ratio of $10^{-1}$, meaning nearly all galaxy mergers with a black hole mass ratio greater than $10^{-1}$ resulted in the orbiting black hole merging with the host.  It is worth noting that all of the mergers were due to N-body effects; without a third interacting black hole, none of the black holes that reached within $r{_c}$ merged with their host.  This conclusion is identical to that reached by the careful treatment by \citet{2018MNRAS.473.3410R} (''Interactions between multiple supermassive black holes in galactic nuclei: a solution to the final parsec problem'').

\begin{figure}
\begin{center}
\includegraphics[width=1.0\columnwidth]{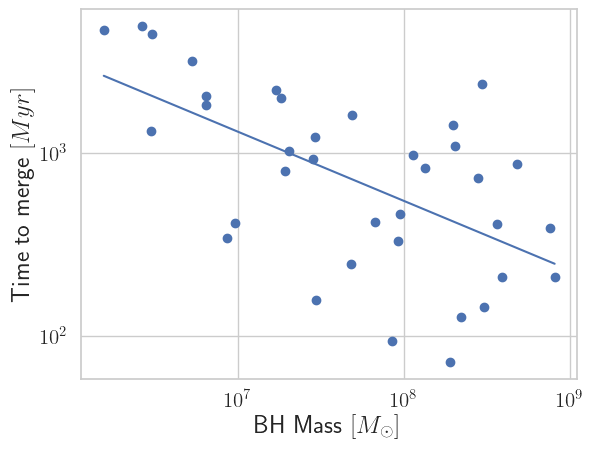}
\caption{Time required for BHs to merge with host SMBH after galaxy merger.  Form for fitted line is $-0.379M_{BH}+5.766$ in log-log space.}
\label{fig:time_to_merge}
\end{center}
\end{figure}

In figure \ref{fig:time_to_merge} we present the time it took for the guest black holes to merge with their host.  As can be seen in the plot, although there is a fair amount of variation, the general trend is a decreasing time to merge as orbiting black hole mass increases.

\subsection{Ejected and Orbiting Black Holes}\label{sec:kicked_orbiting}
Of the 37 orbiting black holes at $z=0$, 18 of them were kicked to higher orbits during traversal of their host galaxy.  If one counts the 13 ejected black holes, then of the 50 SMBHs that had not eventually merged with their host, 32 of them were kicked out of the galactic core as a result of three-body interactions.  If they did not merge, it was not for a lack of effort.

\begin{figure}
\begin{center}
\includegraphics[width=1.0\columnwidth]{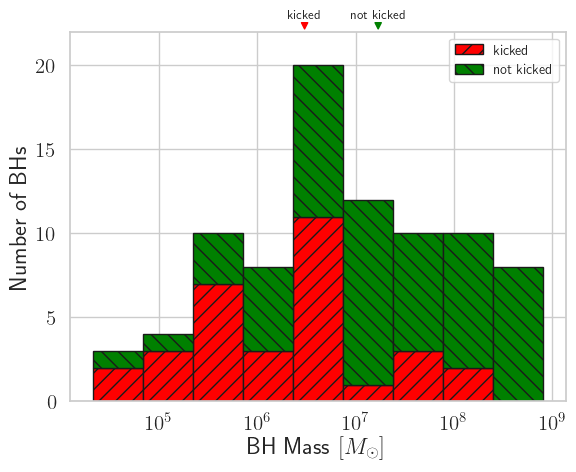}
\caption{Histogram of kicked and not-kicked SMBHs at $z=0$.  Colored arrows along upper x-axis represent median masses of respective sets.  Histograms are stacked.}
\label{fig:kicked_stats}
\end{center}
\end{figure}

Figure \ref{fig:kicked_stats} shows the mass distribution (stacked) of kicked and not-kicked black holes, including those that merged and those that did not merge.  The vast majority of larger SMBHs ($\goa\, 10^7\, M_{\odot}$) were not kicked to higher orbits nor ejected from their galaxies, which at least partly explains why most of the merged SMBHs are in the greater mass range.  In contrast, most of the SMBHs below $10^7\, M_{\odot}$ are kicked (with a few of them being ejected).  In our simulations, no black holes that are kicked to higher orbits ever returned to merge with their central host SMBH.

\begin{figure}
\begin{center}
\includegraphics[width=1.0\columnwidth]{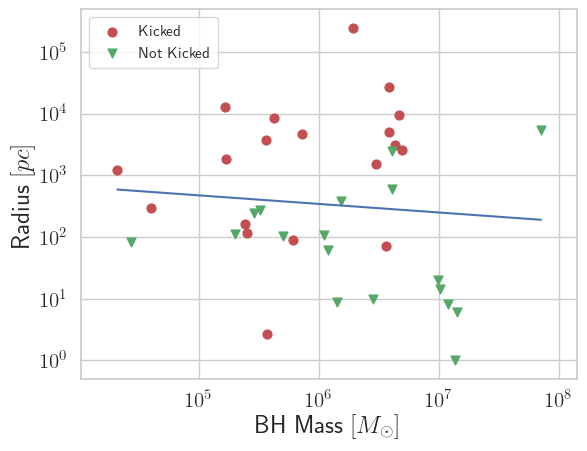}
\caption{Final radii of remaining orbital black holes at $z=0$.  Kicked black holes come back to the center, but due to their highly eccentric nature, they spend most of their time far from the center.  There is a slight inverse relationship between BH mass and final distance from center (solid line).}
\label{fig:mvr}
\end{center}
\end{figure}

Figure \ref{fig:mvr} is a realization of the final radii of the remaining orbiting black holes at $z=0$.  There is a slight inverse relationship between black hole mass and final radius (solid line), but with a wide dispersion.  It is also evident from the plot that all but two of the black holes that had not been kicked by $z=0$ have mostly approached and entered within the galactic core radius, $r_c$ (which never exceeded 1 kpc in any of the galaxies).  The black holes that were kicked come back to the center, but due to their highly eccentric nature, they spend most of their time far from the center.  Additionally, there are only 5 black holes within 10 pc of the center due to N-body effects and the Kozai mechanism.

\begin{figure}
\vspace{20pt}%
\begin{center}
\includegraphics[width=1.0\columnwidth]{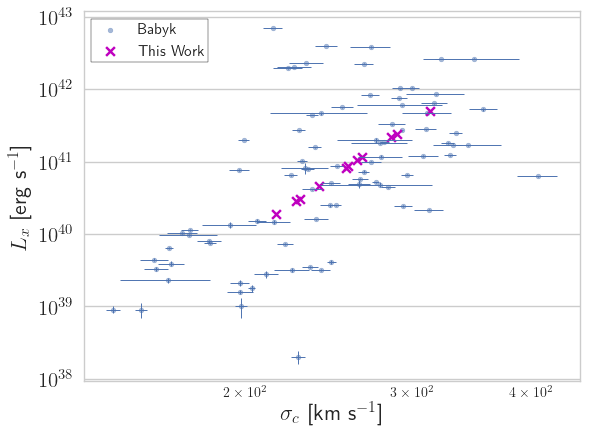}
\caption{X-ray thermal luminosity of hot gas in our galaxies, compared to those from \citet{2018ApJ...857...32B}.}
\label{fig:galgas1}
\end{center}
\end{figure}

\section{Signatures of Orbiting Black Holes}
The large fraction of orbiting black holes found in our simulations motivates a back-of-the-envelope calculation to see if orbiting black holes produce detectable signatures. In the following section, we estimate the X-ray thermal luminosities due to gas accreted by these BHs while orbiting through their host galaxies.

\subsection{Galaxy Gas X-Ray Thermal Luminosity}
First, we construct realistic X-ray thermal luminosities for our host galaxies. We compute the X-ray thermal luminosity of the hot gas in these galaxies, adapting Equation 26 from \citet{2012ApJ...754..125C}. In lieu of summing over discrete particles, we use the beta model of the gas as defined in our equation \ref{beta_model} and integrate:

\begin{equation}
\begin{aligned}
    L_x ={} \frac{1.2 \times 10^{-24}}{(\mu m_p)^2}\left(\frac{kT_x}{1keV}\right)^{1/2}\rho_0^{2}\;4{\pi} \\
    \int_{0}^{\infty}\frac{r^2}{(1+r^2/r_c^2)^{3}}\;dr\text{  erg s}^{-1}.
\end{aligned}
\end{equation}

The integral evaluates to ${\pi}r_c^3/16$.  $m_p$ is the mass of a proton and $\mu$ is the mean molecular weight.  To calculate $\mu$, we use metallicity $Z=2Z_{\odot}=0.0268$, mass fraction of Helium $Y=Y_{\odot}=0.2485$, and mass fraction of Hydrogen $X=0.7247$.  This results in $\mu=0.606$. The resulting X-ray thermal luminosities are shown in Figure \ref{fig:galgas1} and compared against a sample from \citet{2018ApJ...857...32B}.

\subsection{Black Hole X-Ray Thermal Luminosity}
\subsubsection{Assumptions and Method Employed}
Since we model the black holes in this work as point particles for purposes of the simulations, we do not assume any particular structure to them.  Therefore, we adopt the approach and assumptions from \citealt{2018MNRAS.476.1412I} and \citealt{2019MNRAS.486.5377I} in estimating the X-ray luminosities of the remaining black holes left orbiting in our subject galaxies.  Their approach assumes rotating accretion flows in a very low-density environment.  Shear viscous forces transport angular momentum outwards, resulting in an accretion disc that forms at the bottom of the distribution.  However, due to the low density of the gas, it does not cool, and convection limits the net accretion, and subsequent radiation luminosity, of the black hole.  In such a convective infall, the gas finally accreted onto the black hole is much less than the initially infalling gravitationally bound gas.

\subsubsection{Gas Environment Surrounding Black Holes}
In order to model the surrounding environment of the orbiting black holes, we used the data and scaling relations from \citet{2018ApJ...857...32B}, which are based on Chandra X-ray observations of 94 early-type, gas poor galaxies of several morphological types (elliptical, lenticular, SB, BCGs, and cD galaxies).  In particular, we used data from Tables 2 and 3 for temperature in X-ray ($T_x$) and central gas density ($\rho_0$) to model the surrounding gas properties.  Central velocity dispersion ($\sigma_c$) was used as the independent variable.  We estimate the central velocity dispersion in this work based on the potential-density profile in  \citet{2015ApJ...806L..28S} that was also used in modeling the potential in the simulations (cf their equation 11):
\begin{equation}
    \sigma_c^2 = \frac{6GM_{tot}(\frac{\pi^2}{8}-1)}{{\pi}r_h},
\end{equation}
where G is the gravitational constant, $M_{tot}$ is the total galaxy mass (stars + DM), and $r_h$ is the outer halo radius as defined in the Stone-Ostriker profile (cf. equation 1).

\begin{figure}
\vspace{20pt}%
\begin{center}
\includegraphics[width=1.0\columnwidth]{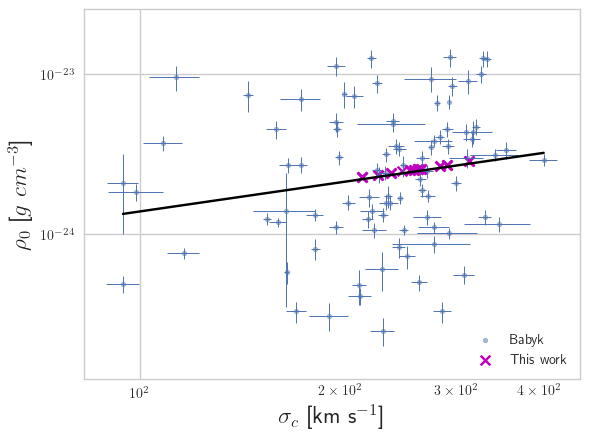}
\caption{Central gas density versus central velocity dispersion.  Purple X's represent central gas densities of the galaxies in this study.  Fit in log space is $\rho_0 = 10^{0.6*\log\left(\frac{\sigma_c}{km\ s^{-1}}\right) - 25} \text{  g cm}^{-3}$.}
\label{fig:rho0_vs_sigma}
\end{center}
\end{figure}

\begin{figure}
\vspace{20pt}%
\begin{center}
\includegraphics[width=1.0\columnwidth]{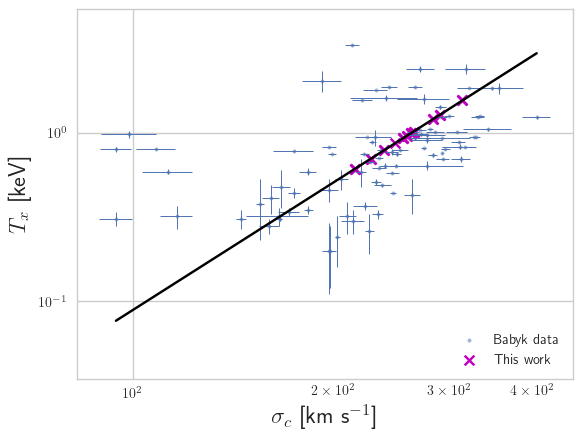}
\caption{X-ray temperature versus central velocity dispersion.  Purple X's represent temperatures of the galaxies in this study.  Fit in log space is $T_x = 10^{2.50*\log\left(\frac{\sigma_c}{km\ s^{-1}}\right) - 6.06} \text{  keV}$.}
\label{fig:Tx_vs_sigma}
\end{center}
\end{figure}

Figures \ref{fig:rho0_vs_sigma} and \ref{fig:Tx_vs_sigma} present the data from \citet{2018ApJ...857...32B}, along with the simple linear fits (in log-log space) and resulting gas properties used for our galaxies:
\begin{subequations} \label{eq:gas_properties}
\begin{eqnarray}
    \rho_0 = 10^{0.6*\log\left(\frac{\sigma_c}{km\ s^{-1}}\right) - 25} \text{  g cm}^{-3},\\
    T_x = 10^{2.50*\log\left(\frac{\sigma_c}{km\ s^{-1}}\right) - 6.06} \text{  keV}.
\end{eqnarray}
\end{subequations}

The dispersion of the data around the fit for $\rho_0$ is at most 1.5-2.0 orders of magnitude.  This leads to a similar variation in the gas density surrounding the black hole, in addition to the ultimate accretion rate of the black hole, due to both having a linear dependence on the central gas density (see Equations \eqref{beta_model} and \eqref{net_accretion} below).

The dispersion of the X-ray temperature data around the fit is approximately 1-1.5 orders of magnitude.  The gas sound speed (see Equation \eqref{gas_sound_speed} below) has a $C_s{\propto}T_x^{0.5}$ dependence on temperature.  However, given the dependence of the accretion rate (see Equation \eqref{net_accretion} below) on gas sound speed of $\dot{M}_{acc}{\propto}C_s^{-3}$, the accretion rate is dependent on temperature to the order of $T_x^{-3/2}$.  This is another source of variation in our estimate of accretion rate.

The local gas density ${\rho}$ surrounding the black hole is obtained from a simple isothermal beta model \citep{1962AJ.....67..471K, 1976A&A....49..137C, 1978A&A....70..677C}:
\begin{equation} \label{beta_model}
    \rho = \frac{\rho_0}{(1+r^2/r_c^2)^{1.5}} \text{  g cm}^{-3},
\end{equation}
where $\rho_0$ is as defined above, r is the black hole's distance from the center of the galaxy, and $r_c$ is the core radius.  Since the orbiting black hole masses are $10^{-5}-10^{-2}M_{tot}$, we assume there is no change in $M_{tot}$ after black holes are ejected.

The sound speed of the gas, $C_s$, is derived from the ideal gas law:
\begin{equation}\label{gas_sound_speed}
    C_s = \sqrt{1.67T_x*1.602\times10^{-16}/m_p} \text{  km s}^{-1},
\end{equation}
where $m_p$ is the proton mass in kilograms.  Since $T_x$ is in keV, the factor of $1.602\times10^{-16}$ is added to convert the temperature to Joules.

\subsubsection{Black Hole Mass Accretion and Luminosity}
At large distances, the black hole mass accretion rate is governed by the Bondi-Hoyle formula (\citealt{1944MNRAS.104..273B} and \citealt{1952MNRAS.112..195B}):
\begin{equation}\label{net_accretion}
    \dot{M}_{B} = \frac{4{\pi}G^2M_\mathrm{BH}^2\rho}{(C_s^2+v^2)^{3/2}} \text{  g}\text{ s}^{-1},
\end{equation}
where $\rho$ is the density of the surrounding gas, $C_s$ is the gas sound speed, and $v$ is the velocity of the black hole.

However, due to convective (i.e., inefficient) accretion interior to the Bondi radius (see \citealt{2018MNRAS.476.1412I} and \citealt{2019MNRAS.486.5377I}), the net accretion to the black hole is often substantially lower than the Bondi-Hoyle accretion.  It can be inferred from the fitting relation given in equation 31 of \cite{2019MNRAS.486.5377I}:
\begin{subequations}
    \begin{equation}\label{inayoshi_accretion}
        \frac{\dot{M}_{\bullet}}{\dot{M}_{Edd}} \simeq 1.5\times10^{-6}T_7^{-4/5}\times\left(\frac{\alpha}{0.01}\right)^{0.37}\left(\frac{\dot{m}_B}{10^{-3}}\right)^{3/5}\left(\frac{f_c}{0.1}\right)^{2/5},
    \end{equation}
    \begin{equation}\label{mdotb}
        \dot{m}_{B} \equiv \dot{M}_{B}/\dot{M}_{Edd},
    \end{equation}
    \begin{equation}\label{edd_acc}
        \dot{M}_{Edd} \equiv \frac{L_{Edd}}{0.1c^2} \text{  g} \text{ s}^{-1},
    \end{equation}
    \begin{equation}\label{edd_lum}
        L_{Edd} \equiv 1.26\times10^{38}\left(\frac{M_{BH}}{M_{\odot}}\right)\;\text{erg}\;\text{s}^{-1}.
    \end{equation}
\end{subequations}
$\dot{M}_{\bullet}$ is the net accretion to the black hole, $T_7$ is the temperature of the ambient gas in units of $10^7\;K$, $\alpha$ is the strength of viscosity (generally accepted to be 0.01; see \citealt{1973A&A....24..337S, 1991ApJ...376..214B, 1995ApJ...445..767M, 1996ApJ...463..656S, 1998RvMP...70....1B, 2004PThPS.155..409S, 2018MNRAS.476.1412I}), and $f_c$ is the conductivity suppression factor (generally accepted to be 0.1; see \citealt{2001ApJ...562L.129N, 2004PhRvL..92d5001M}, and \citealt{2019MNRAS.486.5377I}).  The Eddington accretion rate is $\dot{M}_{Edd}$, $c$ is the speed of light, and $L_{Edd}$ is the Eddington luminosity.

Given the above, we estimate the bolometric luminosity of the black holes as:
\begin{subequations}
    \begin{equation}\label{bol_lum}
        L_{Bol} = \epsilon\dot{M}_{\bullet}c^2,
    \end{equation}
    \begin{equation}\label{rad_eff}
        \begin{split}
        \text{log}\;\epsilon = \begin{cases}
                        -1.0-(0.0162/\dot{m})^{4}\;\;\text{for 0.023}\;\leq\;\dot{m}, \\
                        \sum_{n}{}a_{n}(log\;\dot{m})^{n}\;\;\;\;\;\;\;\;\;\text{for}\;10^{-4} <\;\dot{m}\;<\;0.023, \\
                        \sum_{n}{}b_{n}(log\;\dot{m})^{n}\;\;\;\;\;\;\;\;\;\text{for}\;10^{-8} <\;\dot{m}\;\leq\;10^{-4},
                               \end{cases}
        \end{split}
    \end{equation}
\end{subequations}
where $\dot{m}=\dot{M_{\bullet}}/\dot{M}_{Edd}$, and $\epsilon$ is the radiative efficiency (cf. Equation 13 in \citealt{2019MNRAS.486.5377I}).  The fitted values are $a_0=-0.807$, $a_1=0.27$, $a_n=0$ ($n\geq2$), $b_0=-1.749$, $b_1=-0.267$, $b_2=-0.07492$, and $b_n=0$ ($n\geq3$).

For sources having $\dot{M}_B/\dot{M}_{Edd} > 10^{-3}$, we take $L_x=0.1L_{Bol}$, due to the higher net accretion onto the black hole as a result of cooling of the accretion disc and hence higher gas density.  For those having $\dot{M}_B/\dot{M}_{Edd} < 10^{-3}$, we take $L_x=L_{Bol}$.

\subsubsection{Central Black Hole Influence}
Within the influence radius of the central black hole, which is defined as the radius at which the total mass of the galaxy within that radius (stars + dark matter) equals the mass of the central black hole, the gas density is much higher than otherwise, and equation \eqref{beta_model} cannot be used in the region where the BH Kepler potential dominates.  We start with the equation for hydrostatic equilibrium, assuming isothermality,
\begin{equation} \label{sound_speed}
    C_{s}^2\frac{dln\rho}{dr}= -\frac{GM(r)}{r^2}.
\end{equation}
Inside the sphere of influence, $M(r)$ is constant, and we can integrate Equation \eqref{sound_speed} with respect to r, leading to 
\begin{equation}
    ln\frac{\rho}{\rho_1} = \frac{GM}{C_{s}^2}\left(\frac{1}{r} - \frac{1}{r_1}\right),
\end{equation}
where $\rho$ is the gas density at the location of the orbiting black hole, $\rho_1$ is the density at the radius where the central black hole mass equals the galaxy mass, $r$ is the distance of the orbiting black hole from the center, and $r_1$ is the radius where the central black hole mass equals the galaxy mass.

\begin{figure}
\begin{center}
\includegraphics[width=1.0\columnwidth]{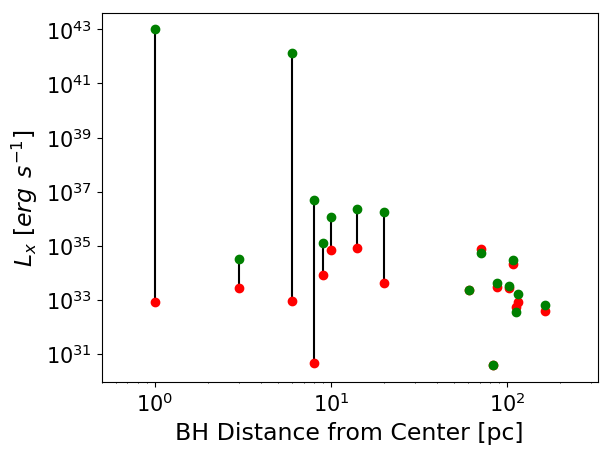}
\caption{Difference in X-ray thermal luminosity of orbiting black holes within the influence radius of the central black hole.  For each vertical line, red (lower) dots are the originally calculated luminosity, not taking the CBH influence into account.  Green (upper) dots are the final luminosities, taking the central BH influence into account.}
\label{fig:wui}
\end{center}
\end{figure}

Figure \ref{fig:wui} shows the difference in luminosity when taking into account the influence of the central black hole for the 17 orbiting black holes that were within the influence radius.  In all cases, the final luminosity is higher than the original luminosity.

\subsubsection{Final X-ray Luminosities of orbiting Black Holes, and Expected Detection Spectra}
\begin{figure}
\begin{center}
\includegraphics[width=1.0\columnwidth]{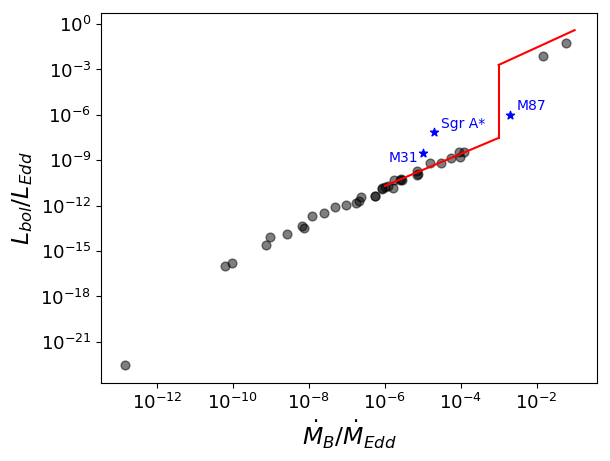}
\caption{Approximation to Figure 12 from \citealt{2019MNRAS.486.5377I}. Black "o"s represent black holes in this study.  We also include observational results for Sgr A*, and BHs in M31 and M87.}
\label{fig:loobhs}
\end{center}
\end{figure}

Figure \ref{fig:loobhs} is an approximation to Figure 12 from \cite{2019MNRAS.486.5377I}, along with the final $L_{Bol}/L_{Edd}$ ratio of the orbiting black holes from this study, although extended below $10^{-6}$ $\dot{M}_{B}/\dot{M}_{Edd}$ for the very low accretion-rate sources.  Indeed, such low luminosity sources are common, with some estimates of $L_{Bol}/L_{Edd}$ as low as $10^{-10}$ ( e.g., \citealt{2003ApJ...598..301Y}, \citealt{2004ApJ...613..322Q}, \citealt{2008ARA&A..46..475H}, \citealt{2009ApJ...699..626H}, \citealt{2018MNRAS.476.1412I}).

\begin{figure}
\begin{center}
\includegraphics[width=1.0\columnwidth]{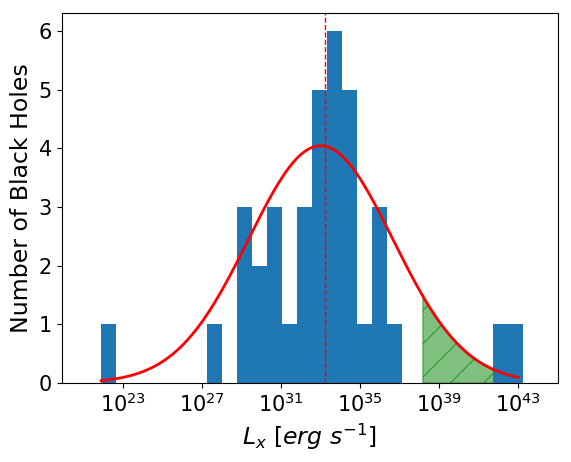}
\caption{X-ray thermal luminosity distribution of orbiting BHs at $z=0$.  Green region under curve represents approximate range of detectable orbiting black holes, with the lower limit being approximately $10^{38}\;\text{erg}\;\text{s}^{-1}$ (\citet{2018ApJ...862...73L}). The dashed, red vertical line is the median X-ray thermal luminosity of our sources ($1.0\times10^{33}\;\text{erg}\;\text{s}^{-1}$).}
\label{fig:ldobhs}
\end{center}
\end{figure}

Figure \ref{fig:ldobhs} represents a histogram of the final X-ray luminosities of the black holes, overlaid by a lognormal distribution, obtained with a mean $\mu=1.0\times10^{33}\,\text{erg s}^{-1}$ and standard deviation $\sigma=4.5\times10^{3}\,\text{erg s}^{-1}$.

The orbiting black holes should be detectable via their anomalous soft X-ray spectra. Sgr A*, the black hole at the center of the Milky Way Galaxy, has been observed to fit a power law spectrum with photon index $\Gamma\sim2.7^{+1.3}_{-1.9}$ \citet{2003ApJ...591..891B}.  Conversely low-mass X-ray binaries (LMXRB) have been fit to a power law spectrum with $\Gamma\sim1.5-2.0$ (\citet{2018Natur.556...70H}, and we expect our orbiting sources to be more similar to Sgr A* than to normal LMXRBs.

\citet{guo2020hunting} have shown that there is an additional way of detecting the predicted orbiting massive black holes. They compute that the spectra have a peak at the millimeter wave band, and should enable detectors such as the Atacama Large Millimeter/submillimeter Array (ALMA) and upcoming next generation Very Large Array (ngVLA) to find orbiting black holes with masses down to approximately $M_{\bullet}\simeq2\times10^7 M_{\odot}$ in nearby galaxies such as M87, and $M_{\bullet}\simeq10^5 M_{\odot}$ in the Milky Way.

\section{Conclusions}\label{sec:conclusions}
We performed N-body simulations on 11 galaxies ranging in mass between $10^{12}$ M$_{\sun}$\;<\;M\;<\;3$\times$ $10^{13}$ M$_{\sun}$, from $4\;\geq\;z\;\geq\;0$.  A total of 86 orbiting black holes were followed in our simulations, resulting in 42$\%$ mergers with their host galaxy's resident black hole, while another 43$\%$ remained in orbit at $z=0$.  Those remaining orbiting black holes had X-ray luminosities in the range of $10^{28}-10^{43}\mathrm{erg}~\mathrm{s}^{-1}$.  The median luminosity of $10^{33}\mathrm{erg}~\mathrm{s}^{-1}$ would be undetectable with present instruments, although the top $5\%$ most luminous from this set could be detectable.  If detected, these orbiting massive black holes would constitute a new category of X-ray source.

\section{Acknowledgements}
The authors would like to thank Andrea Kulier and Claire Lackner for providing merger tree data, and Seppo Mikkola for making his \textsc{AR-Chain} code available. We also thank Renyue Cen, Pieter van Dokkum, Charles Hailey, Zoltan Haiman, Kohei Inayoshi, Sean McWilliams, Taeho Ryu and Nick Stone for useful discussions and insights.  AHWK acknowledges support by NASA through Hubble Fellowship grant HST-HF-51323.01-A awarded by the Space Telescope Science Institute, which is operated by the Association of Universities for Research in Astronomy, Inc., for NASA, under contract NAS 5-26555.




\bibliographystyle{mnras}
\bibliography{biblio}

\bsp	
\label{lastpage}
\end{document}